\documentclass[sigconf,authorversion]{acmart}

\usepackage{fancyhdr}
\AtBeginDocument{%
    \addtolength{\footskip}{2.0\baselineskip}%
    \fancyfoot[C]{\textit{{Preprint --- do not distribute.}}}%
}

\usepackage[utf8]{inputenc}
\usepackage[english]{babel}

\DeclareUnicodeCharacter{0229}{\c{e}}
\usepackage{soul}
\usepackage{framed}
\usepackage[strict]{changepage}
\usepackage{fancybox}

\usepackage{color}
\definecolor{colorbar}{HTML}{9f5a7f}
\definecolor{formalshade}{HTML}{eaeaf2} 

\newenvironment{xdefinition}{%
	\interlinepenalty=10000
	\MakeFramed{\advance\hsize-\width\FrameRestore}%
	\noindent\hspace{-4.55pt}
	\begin{adjustwidth}{}{4pt}%
		\vspace{1pt}%
	}
	{%
	\vspace{2pt}\end{adjustwidth}\endMakeFramed%

}

\usepackage{enumitem}
\usepackage{tabularx}
\usepackage{makecell}
\usepackage{multirow}
\usepackage{graphicx}
\usepackage{amsfonts}

\renewcommand{\arraystretch}{1.1}
\definecolor{formalshade}{HTML}{f7f7f7} 
\graphicspath{ {./figures/} }

\AtBeginDocument{%
 }

\setcopyright{acmcopyright}
\copyrightyear{2025}
\acmYear{2025}
\acmDOI{}

\acmConference[EASE 2025]{The 29th International Conference on Evaluation and Assessment in Software Engineering}{17–20 June, 2025}{Istanbul, Türkiye}

\acmPrice{15.00}
\acmISBN{978-1-4503-XXXX-X/18/06}

\begin{document}

%Maße des Layouts:
%paperwidth in cm: \printinunitsof{cm}\prntlen{\paperwidth}
%17.7917cm

%columnwidth in cm: \printinunitsof{cm}\prntlen{\columnsep}
%0.84338cm

\title[Modeling Communication Perception in Development Teams Using Monte Carlo Methods]{Modeling Communication Perception in Development Teams Using Monte Carlo Methods}

\author{Marc Herrmann}
\orcid{0000-0002-3951-3300}
\affiliation{%
  \institution{Leibniz University Hannover}
  \department{Software Engineering Group}
  \city{Hannover}
  \country{Germany}
}
\email{marc.herrmann@inf.uni-hannover.de}

\author{Martin Obaidi}
\orcid{0000-0001-9217-3934}
\affiliation{%
  \institution{Leibniz University Hannover}
  \department{Software Engineering Group}
  \city{Hannover}
  \country{Germany}
}
\email{martin.obaidi@inf.uni-hannover.de}

\author{Jil Klünder}
\orcid{0000-0001-7674-2930}
\affiliation{%
  \institution{Leibniz University Hannover}
  \department{Software Engineering Group}
  \city{Hannover}
  \country{Germany}
}
\email{jil.kluender@inf.uni-hannover.de}

\renewcommand{\shortauthors}{Herrmann et al.}

\begin{abstract}
Software development is a collaborative task involving diverse development teams, where toxic communication can negatively impact team mood and project success. Mood surveys enable the early detection of underlying tensions or dissatisfaction within development teams, allowing communication issues to be addressed before they escalate, fostering a positive and productive work environment. The mood can be surveyed indirectly by analyzing the text-based communication of the team. However, emotional subjectivity leads to varying sentiment interpretations across team members; a statement perceived neutrally by one developer might be seen as problematic by another developer with a different conversational culture. Early identification of perception volatility can help prevent misunderstandings and enhance team morale while safeguarding the project.

This paper analyzes the diversity of perceptions within arbitrary development teams and determines how many team members should report their sentiment to accurately reflect the team's mood. Through a Monte Carlo experiment involving 45 developers, we present a preliminary mathematical model to calculate the minimum agreement among a subset of developers based on the whole team's agreement. This model can guide leadership in mood assessment, demonstrating that omitting even a single member in an average-sized 7-member team can misrepresent the overall mood. Therefore, including all developers in mood surveying is recommended to ensure a reliable evaluation of the team's mood.
\end{abstract}

% CCS XML
\begin{CCSXML}
<ccs2012>
<concept>
<concept_id>10011007.10011074.10011134</concept_id>
<concept_desc>Software and its engineering~Collaboration in software development</concept_desc>
<concept_significance>500</concept_significance>
</concept>
<concept>
<concept_id>10011007.10011074.10011134.10011135</concept_id>
<concept_desc>Software and its engineering~Programming teams</concept_desc>
<concept_significance>500</concept_significance>
</concept>
<concept>
<concept_id>10010147.10010341</concept_id>
<concept_desc>Computing methodologies~Modeling and simulation</concept_desc>
<concept_significance>300</concept_significance>
</concept>
</ccs2012>
\end{CCSXML}

\ccsdesc[500]{Software and its engineering~Collaboration in software development}
\ccsdesc[500]{Software and its engineering~Programming teams}
\ccsdesc[300]{Computing methodologies~Modeling and simulation}
\ccsdesc[100]{Mathematics Subject Classification (2020)~68N30; 65C05; 62J02}

\keywords{Software engineering, communication, collaboration, mood analysis, sentiment analysis, perception, agreement, monte carlo experiment, nonlinear regression}

\received{31 January 2025}
\received[revised]{27 April 2025}
\received[accepted]{21 March 2025}


\begingroup
\def\UrlFont{\normalsize} %smaller email
\mathchardef\UrlBreakPenalty=10000
\maketitle
\endgroup

\fancypagestyle{plain}{%
  \renewcommand{\headrulewidth}{0pt}%
  \fancyhf{}%
  \fancyfoot[C]{\textit{{Preprint --- do not distribute.}}}%
}
\thispagestyle{plain}

\section{Introduction}
\label{sec:introduction}

Software projects require teamwork and social interactions in order to be successfully completed~\cite{kraut1995coordination,zaman2019socialskills}. Open and respectful communication fosters a positive work environment. It encourages trust, collaboration, and a sense of belonging within the team\cite{awotunde2020colab}. Timely and clear communication helps address and resolve conflicts, misunderstandings, and differences in opinions before they escalate. When communication is perceived as clear, open, and effective, team members feel more connected, supported, and engaged~\cite{demarco2013peopleware}. This can lead to better collaboration, higher morale, and a more cohesive work environment~\cite{schneider2018positive}. That is, if the team communicates in a friendly way and is generally in a good mood, the project is more likely to be finished successfully~\cite{graziotin2014happy,graziotin2015you,schneider2018positive}.

Poor communication on the other hand can lead to misunderstandings, frustration, and a lack of clarity about responsibilities. This can result in duplicated efforts, missed deadlines, and lower quality outcomes~\cite{stapel2014information-flow}. Negative communication can breed a toxic work environment in software projects~\cite{miller2022toxicity}, where conflicts go unresolved, and team members feel undervalued or unsupported, which can lead to far-reaching consequences such as burnout~\cite{tulili2022burnout}. Such symptoms are sometimes described as community smells~\cite{caballero2023community-smells}, i.e., suboptimal social patterns and organizational structures within a development team that can negatively impact collaboration, communication, and ultimately the software quality.

To mitigate these problems, project leaders can survey the mood of the team by  interviews with individual team members. However, this may be ineffective if the mood of the developers surveyed does not reflect that of the rest. Another technique to mitigate these problems is the application of sentiment analysis for software engineering (hereinafter SA4SE), where a software analyzes the text-based communication within collaborative software development environments and determines whether a statement's sentiment polarity is \textit{positive}, \textit{negative}, or \textit{neutral}~\cite{medhat2014sentimentanalysis,obaidi2021slr}. This way, a project leader gathers information on the amount of negative communication, assuming that this is an indicator for problems or points to the need for interventions~\cite{herrmann2021meeting, sentianalyzerreport2022, schroth2022potential, specht2024hcse}.

However, a recent study among software developers~\cite{herrmann2022subjectivity} has shown that the perception of developers differs remarkably from the predefined sentiment polarities in SA4SE-datasets. While some developers perceive a statement as \textit{positive}, others perceive it as \textit{neutral} or even as \textit{negative}~\cite{herrmann2025different-perceptions}. The significant variability in perceptions among software developers prompts an inquiry into whether excessive variability exists within arbitrary software development teams and how many team members are necessary to accurately represent the overall mood trend, e.g., for surveying the mood of selected developers or calibrating a sentiment analysis approach to the teams' communication.

Despite its critical nature, communication within software development teams often suffers due to various factors such as diversity, varying levels of experience and expertise, and most importantly differing perceptions of communication~\cite{herrmann2025different-perceptions} among team members. However, hardly any study examines the differences in the perceptions of developers, let alone attempt to model them formally.

In this paper, we want to investigate how volatile the agreement towards perception of software project communication among teams of developers is. Further, we want to find out how many developers from a larger team need to be surveyed in order for project managers to elicit the mood of the team. We perform a Monte Carlo experiment~\cite{paxton2001monte-carlo} based on a developer sentiment survey study conducted in 2021~\cite{sentisurvey-zenodo}. We modeled the perceptions of 45 software developers with varying levels of experience and skills with respect to their agreement. We provide the following key contributions:

\begin{itemize}[leftmargin=15.5pt]
    \setlength\itemsep{1mm}
    \item[\textbf{(1)}] We compute numerous arbitrary draws to calculate deviations to the agreement of a group of developers by bootstrapping a random subset of developers. These deviations can then be used to determine the volatility and the expected intervals of agreement of perceived communication in software projects.

    \item[\textbf{(2)}] Furthermore, we provide a preliminary mathematical model to calculate the minimum agreement of a subset of raters when compared to the agreement of the whole set. 

    \item[\textbf{(3)}] Lastly, we recommend best practices about the number of sufficient raters needed to match the perception of an average sized development team based on our findings, which imply how many developers should be surveyed for mood analysis.
\end{itemize}

\textit{Outline.} The rest of the paper is structured as follows: In Section~\ref{sec:background}, we highlight related research and background details. Section~\ref{sec:research} explains the design of the experiment, including our data and methodology. The results are presented in Section~\ref{sec:results} and discussed in Section~\ref{sec:discussion}. The paper is summarized in Section~\ref{sec:conclusion}.

\section{Background and Related Work}\label{sec:background}
This background section reviews key research that reveals the intricate interplay of cultural, demographic, and personality factors that influence emotional perception, and further the methodological factors in mood and sentiment analysis within software engineering. They advocate for a multi-faceted approach that considers these variables to accurately interpret team emotions, thereby potentially enhancing collaborative outcomes in development projects.

\subsection{Diversity of Sentiment Perception}
Understanding the nuances of sentiment perception in communication is crucial for accurately interpreting the affective states of individuals within development teams. Previous research has explored various factors that influence how sentiments are perceived and expressed, highlighting the importance of cultural and demographic considerations. Heise~\cite{heise2014culturalsentiments} underscores the significant impact of cultural background on the perception of emotional meanings in communication. By examining sentiment perception across 17 different cultures, Heise identified secularism and the historical influence of culture during colonization as key factors that affect how emotional content is interpreted, highlighting the importance of acknowledging cultural diversity when analyzing communication within global or multicultural development teams.

The impact of ethnicity and race on communication styles was further explored by Preoţiuc-Pietro and Ungar~\cite{preotiuc-pietro-ungar-2018-user}, who studied an extensive dataset of over 5,000,000 tweets to develop a classifier capable of predicting users' race and ethnicity based on their communication styles. Achieving a high AUC-value of 0.88 for major racial and ethnic groups, their findings suggest that subtle cues in language use can serve as indicators of ethnic and racial identity. Demographic factors such as age and gender also play a significant in shaping communication expression and perception, as demonstrated by Rehan et al.~\cite{rehan2023age-gender-sentiments}. In their analysis of online product reviews, they discovered that younger individuals tend to exhibit informal language and more extreme emotional expressions, while older individuals tend to adopt a more moderate tone. Additionally, gender differences were observed in how emotions were expressed and the focus of communication, underscoring the need to account for these demographics for accurate sentiment interpretation.

Herrmann et al.~\cite{herrmann2025different-perceptions} investigate how software developers perceive sentiments in project communications by examining the diverse perceptions within software development teams. The authors perform a hierarchical cluster analysis on 94 developers, identifying two distinct groups based on their sentiment perceptions toward statements derived from software project communications. Despite the significant differences in sentiment perception identified for about 65\% of the statements, demographic characteristics such as age and experience did not significantly differ between these groups. The authors suggest that sentiment perception among developers is influenced by factors beyond traditional demographics, potentially involving cultural backgrounds, human values~\cite{schwartz1992values}, and personality traits~\cite{madison2021ocean}.

Despite these differences in perception of communication, studies on community dynamics based on simulation methods so far only concentrate on investigating the global software engineering community at large~\cite{tamburri2012monte-carlo-se}.

\subsection{Interrater Agreement of Developers}
Emam~\cite{emam1999benchmarking} researched the reliability of software process assessments, highlighting the frequent use of Cohen's $\kappa$~\cite{cohen1960agreement} statistic to measure interrater agreement. Historically, benchmarks for interpreting $\kappa$ values have been drawn from social science and medical fields, but their applicability in software assessments is difficult due to contextual variations. This underscores the need for a domain-specific benchmark that accurately evaluates the reliability of new assessments and fosters improvements in assessment methods. Furthermore,  the importance of exploring how assessment reliability impacts decision-making was identified, which could lead to alternative benchmarks. In alignment with this, Jung~\cite{jung2003evaluating} emphasized the importance of measurement reliability in validating software process assessment outcomes, focusing on methods for evaluating interrater reliability. The use of the $\kappa$ coefficient was advocated, alongside the index of observed agreement, to provide a comprehensive view of assessment reliability.

\subsection{Sentiment Analysis in Software Engineering}
Recent research has frequently analyzed the potential of sentiment analysis in the context of software projects~\cite{obaidiSentiSMS22, novielli2023emotion-analysis}. Most papers focus on the development of new tools, the improvement of existing tools, and their application~\cite{obaidiSentiSMS22}. Lin et al.~\cite{lin2022slr} identified limitations of general sentiment analysis tools in software contexts and underscored the need for customized approaches. Further, several studies on sentiment analysis in software engineering emphasize the significance of interpreting emotional expressions accurately, given their potential impact on collaborative development environments. 

Murgia et al.~\cite{Murgia.2014} conducted a study with 16 participants, revealing that while there is notable agreement on neutral sentiments, consensus on emotions such as love, joy, and sadness was less consistent. They observed that the addition of more than two raters did not enhance the interrater agreement, which suggests that no more than two human raters are needed in evaluating sentiment polarities like positive, neutral, and negative in the software engineering domain. Further challenges were noted by Jongeling et al.~\cite{jongeling2017negative}, who found limited agreement between general-purpose sentiment analysis tools and human annotations in software engineering contexts. Their work underscores the necessity for sentiment analysis tools specifically tailored for this domain (SA4SE-tools) and highlights the of subjectivity in affecting consensus.

Adding to this discourse, Novielli et al.~\cite{novielli2018stackgold, novielli2020githubgold} explored the application of emotion classification frameworks in sentiment analysis by assembling datasets from Stack Overflow and GitHub. They applied Shaver et al.'s six basic emotions model~\cite{shaver1987emotion}. Their work demonstrates the importance of employing consistent emotional categories across various datasets to capture the emotional nuances present in developer communications. Similarly, Ding et al.~\cite{ding2018problems} addressed sentiment polarities on GitHub communication, resolving polarity disagreements with additional rater input, which aligns with Murgia et al.'s~\cite{Murgia.2014} findings on rater consensus limitations.

Ortu et al.~\cite{ortu2016jira} and Islam and Zibran~\cite{islam18deva} further enriched this field by curating Jira datasets with distinct emotion models, such as the Parrott model~\cite{parrott2001emotions} and Russell and Mehrabian's model~\cite{russell97emotionmodel}, respectively. These works illustrate the variability and application of different emotional models in sentiment analysis for software engineering, acknowledging that diverse tools and models can yield different insights when dealing with developer communications. Similarly, Uddin and Khomh~\cite{uddin2021problems} and Lin et al.~\cite{lin18sentiment} contributed by focusing on Stack Overflow discussions and constructed datasets emphasizing opinion polarity scales, thus broadening the scope of sentiment analysis across different software engineering communication platforms.

Imtiaz et al.~\cite{imtiaz18problems} and Ahmed et al.~\cite{ahmed2017senticr} focused on GitHub and code review datasets, respectively, highlighting methodological approaches to labeling sentiment polarities and binary-class sentiments in open-source project communications. Finally, the APP reviews dataset, as assembled by Villarroel et al.~\cite{villarroel16appreviews} and subsequently labeled by Lin et al.~\cite{lin18sentiment}, serves as a case study in applying sentiment analysis techniques across diverse review types, thereby illustrating the adaptability and challenges of these methodologies in different contexts.

Herrmann et al.~\cite{herrmann2022subjectivity} examined two established SA4SE-datasets~\cite{lin18sentiment,novielli2020githubgold} by comparing them to the perceptions of a group of software developers ($n = 94$). They found flaws in the ability of the predefined labels from the scientific authors in both datasets when comparing them to the median perception of the developers. This was especially true for the perceptions of each individual developer: while some participants achieved good agreement with the authors' predefined sentiment polarities, other participants showed considerable disagreement with the authors~\cite{herrmann2022subjectivity}. However, the results also showed that the use of emotion frameworks during the data annotation improves the agreement to the median perception of developers for the resulting dataset~\cite{herrmann2022subjectivity}. These insights also raise the question of how the mood of a software development team can be analyzed if there is large variation in the perception of sentiments within individual developers. 

\section{Research Design}
\label{sec:research}
Inspired by the related work, we aim to explore how many team members should report their sentiment to accurately reflect a development team’s mood. In the following subsections, we present our research objective and the research questions, as well as the structure of the case survey, followed by the statistical and mathematical methodology for answering our research questions.

\subsection{Research Objective and Research Questions}
Our main research objective is \textit{to analyze how agreement varies with small samples of developers} and \textit{to examine how many persons are needed to meet the perception of a development team} from a mathematical viewpoint. In addition, we want to derive a model that estimates the minimum agreement of a sample of developers to the agreement of the whole development team. To achieve these goals, we pose the following research questions:

\begin{itemize}[leftmargin=25pt]
    \setlength\itemsep{1mm}
    \item[\textbf{RQ1:}] \textit{How does the agreement of sentiment towards software project communication among a subset of developers change with decreasing subset size?}
    \item[\textbf{RQ2:}] \textit{How can the minimum agreement for small developer subsets be mathematically modeled based on the overall sentiment agreement of the entire group?}
    \item[\textbf{RQ3:}] \textit{What is the minimum number of developers required to achieve the same level of sentiment agreement as an average-sized software development team?}
\end{itemize}

\subsection{Data Basis}
We use an existing data basis that contains information on how 180 software developers with varying levels of experience perceive 100 statements from two popular SA4SE-datasets. The raw survey responses are available via \href{https://doi.org/10.5281/zenodo.6611728}{Zenodo}~\cite{sentisurvey-zenodo}. Therefore, our results cannot be generalized for other groups of persons, but provide insights on the diversity of a software development team. Details on the survey development, including the origin of the statements to be labeled, and the data collection process are provided by Herrmann et al.~\cite{herrmann2022subjectivity}. In the following, we only describe the survey structure, the altered data pre-processing and the data analysis. 

\subsubsection{Survey Structure}
The questionnaire consisted of five sections (number of questions in brackets): Demographics (4), affiliation to computer science (2), programming experience (5), labeling (1 repeated for 100 statements), and criteria (1). Initially, demographic data such as age, gender, native language, and frequency of communication in English were collected. The next block of questions inquired about participants' identification as developers and their current professional status. Afterward, questions focused on programming skills, experience in professional work environments, and teamwork. Following that, the survey included a block of 100 statements. These statements comprised 48 distinct statements from the gold standard dataset~\cite{novielli2020githubgold} and 48 statements from the ad hoc labeled dataset~\cite{lin18sentiment}, with 4 duplicates. We included 4 random statements (2 \textit{positive}, 1 \textit{neutral}, and 1 \textit{negative}) from the GitHub gold standard dataset~\cite{novielli2020githubgold} twice, resulting in a total of 100 statements. This was done to assess the consistency of participants' labeling, specifically to determine if they assigned different labels to any of the duplicate statements during the survey. Participants were requested to classify each statement into one of three polarity classes: \textit{positive}, \textit{neutral}, or \textit{negative}, without any specific guidelines provided. Further details regarding the selection of these statements are provided in the following subsection. The 100 statements in the survey were chosen randomly, resulting in an approximate one-third split for each of the three sentiment polarity classes, with 32 \textit{positive}, 32 \textit{neutral}, and 32 \textit{negative} statements (plus 4 duplicates). It should be noted that participants were not informed about the equal distribution of statements among the polarity classes to avoid bias. The statements were organized into blocks of ten, and both the blocks and the statements within each block were randomly ordered. The last question inquired about participants' approach to selecting the labels, offering predefined answer options such as focusing on statement content or tone, as well as providing a free-text answer.

\subsection{Data Pre-Processing}
\label{subsec:preprocess}
Our data analysis is based on 180 data points~\cite{sentisurvey-zenodo}, each containing 127 parameters, with some resulting from optional questions.

To answer the research questions, we applied two conditions when defining the subset of the 180 data points being suitable for the analysis presented in this paper. First, we removed 17 data points that answered either question on programming experience or being a computer scientist (\textit{``Would you identify yourself as a computer scientist (e.g., computer science student, developer, etc.)?\hphantom{}''} or \textit{``Do you have any experience with programming (e.g., programming a software, website, an app, etc.)?\hphantom{}''}) with a ``No'', as our target group were persons who are potential members of a software project team. As a second criterion, we only included data points where participants had annotated all 100 statements with their perceived sentiment polarity, because the interrater agreement (i.e., Fleiss'~$\kappa$~\cite{fleiss1971agreement}) can only be calculated if each rater has labeled each statement, removing a further 118 data points. Applying these criteria led to a final dataset consisting of 45 data points that were used for the further analysis procedures of our research questions. 

\subsection{Used Variables}
This paper is mainly based on two types of variables. In the dataset, we have 100 statements $s_i,\, i \in \{1, 2, \ldots, 100\}$ that were annotated by each of the 45 participants $p_j,\, j \in \{1, 2, \ldots, 45\}$. Each participant $p_j$ assigned a label $l \in \{\textit{positive, negative, neutral}\}$ to each statement $s_i$, so we have a total of 4500 labels $l(s_i, p_j)$ where $1 \leq i \leq 100,\; 1 \leq j \leq 45$ that are considered in our analysis. As we are interested in deviations of agreement, we calculate the interrater agreement using Fleiss'~$\kappa$ for randomly bootstrapped subsets of all 45 participants across all statements $s_i$. The agreement for all 45 participants will always be the same value $\hat{\kappa}$, while the agreement of a random subset $\mathcal{P}$ depends on the sentiment labels $l(s_i, p_j)$ of the randomly selected participants $p_j \in \mathcal{P}$.

\subsection{Data Analysis}
\label{subsec:analysis}
We analyzed the collected and pre-processed data using Python with various packages for scientific research including \textit{pandas}~\cite{mckinney2010pandas}, \textit{NumPy}~\cite{harris2020numpy}, \textit{Matplotlib}~\cite{hunter2007matplotlib}, \textit{Scikit-learn}~\cite{pedregosa2011scikit}, and \textit{SciPy}~\cite{virtanen2020scipy}. In the following sections, we present the detailed analysis procedures for each research question. 

\subsubsection{Analysis Procedures for RQ1}
\label{subsec:rq1} 
For RQ1, we wanted to analyze how much the agreement of developers varies when only small subsets of them are considered. If we wanted to analyze all possible subsets of our 45 participants in terms of deviations from $\hat{\kappa}$, that would have led to $44! \approx 2.7 \cdot 10^{54}$ combinatory possibilities. Therefore, we instead chose to conduct a Monte Carlo experiment~\cite{paxton2001monte-carlo} to cover a representative range of the combinatorial possibilities. The general setup of our Monte Carlo experiment, which we used to examine the progression of the interrater agreement with an increasing numbers of raters $n$ out of a group of $k$ total persons, is provided in Table~\ref{table:mc}.

\begin{table}[htb]
\setlength{\tabcolsep}{3pt}
\caption{Algorithm used for the Monte Carlo experiment.} 
\begin{center}
\begin{tabularx}{\columnwidth}{lX}
\Xhline{2\arrayrulewidth}\addlinespace
    \textbf{Step 1:} & Select a subset $\mathcal{P} \subseteq \{p_1, p_2, \ldots, p_{45}\}$ of $\vert \mathcal{P} \vert = k$ participants such that $3 \leq k \leq 45$.\\ \addlinespace
    \textbf{Step 2:} & Generate a randomly ordered sequence $\mathcal{S}$ of all $k$ participants $p_j \in \mathcal{P}$.\\ \addlinespace
    \textbf{Step 3:} & Calculate $\kappa_2, \kappa_3, \ldots ,\kappa_{k-1}, \kappa_{k}$, where $\kappa_n$ denotes the interrater agreement of the first $2 \leq n \leq k$ participants $p_j \in \mathcal{S}$. The final interrater agreement $\kappa_{k}$ of all $k$ participants $p_j \in \mathcal{P}$ independently of $\mathcal{S}$ always takes on the same value $\hat{\kappa}$.\\ \addlinespace
    \textbf{Step 4:} & Repeat Steps 2, 3 for $m - 1$ times to generate a total of $m$ random progressions of the interrater agreement with an increasing number of 2, \ldots, $k$ participants.\\
\addlinespace\Xhline{2\arrayrulewidth}
\end{tabularx} 
\label{table:mc}
\end{center}
\end{table}

We generalized this algorithm for final group sizes of $3 \leq k \leq 45$ participants, so that we can investigate the behavior for more appropriately sized development teams, which we needed to answer RQ2 and RQ3. Therefore, Step~1 in Table~\ref{table:mc} would select an arbitrary team of $k$ developers on which we perform the Monte Carlo experiment. By repeating the random sampling for $m$ total times we apply the law of large numbers~\cite{dekking2005law-of-large-numbers} thus ensuring valid statistical interference of the experiment. To present the results of our Monte Carlo experiment comprehensibly, we visualized the deviations of agreement using a violin plot, where each violin represents the distribution of all $m$ agreement values $\kappa_n$ for a subset of $n$ participants.     

\subsubsection{Analysis Procedures for RQ2}
\label{subsec:rq2}
To answer RQ2, we want to describe the minimum agreement boundary for subsets of developers with a mathematical model based upon given knowledge about the whole set. Our observations resulting from the analysis procedures for \textit{RQ1} showed that by picking a smaller number of developers out of a bigger group, one could get either lucky by picking a few raters that agree more upon each other than the bigger group, but also unlucky by picking a few raters that substantially disagree with each other. The latter could lead to really inconsistent datasets when assigning disagreeing raters with the annotation task. To quantify how bad the possible disagreement can be, we determined the minimum value of agreement (i.e., $\min \kappa_n$) for each $n$ raters resulting from $m$ repeated cycles as defined within algorithm~\ref{table:mc}. We contemplated these minimum values in correspondence to the $n$ raters individually and tried fitting a mathematical model to the values using non-linear multiple regression analysis in \textit{SciPy}~\cite{virtanen2020scipy}. We started with a regression model $f(n, a, b, \ldots)$ using $n$ as a parameter of $f$ as we know from \textit{RQ1} that $n$ effects $\min \kappa_n$. Additionally, we used as many regressors ($a, b, \ldots$) to be determined by the regression as needed to fit the data. We repeated many runs of the Monte Carlo experiment itself to find out which regressors behave as constant values. Furthermore, we analyzed which regressors were influenced by changes of the team size $k$ and  the final agreement value $\hat{\kappa}$. Therefore, we could iteratively replace regressors of the model with their assumed real values, thus increasing the accuracy of the model for the remaining regressors. To evaluate the fit of the model $f$ to the true values $\min \kappa_n$, we computed the coefficient of determination $R^2$~\cite{nagelkerke1991r2}. 

\subsubsection{Analysis Procedures for RQ3}
\label{subsec:rq3}
For RQ3, we wanted to know how many developers of an average sized software development team can be neglected while maintaining an agreement matching the whole team reasonably close. To answer RQ3, we firstly had to set $k$ to fit the number of persons in an average sized software development team. For demonstration, we consider $k = 7$ to be a commonly sized agile development team nowadays, as most of the early Scrum and XP books all suggest a team size number of $7 \pm 2$~\cite{sutherland2019scrum}, applying Miller’s Number~\cite{miller1956magical}. We therefore generated data using our Monte Carlo experiment with $k = 7$. The resulting randomly sampled values of $\kappa_n$ are normally distributed according to the central limit theorem~\cite{laplace1810clt}. Using the empirical rule, we can therefore calculate interval estimates for an observation of $\kappa_n$, based on the mean $\mu_n$ and standard deviation $\sigma_n$ of our data as follows:
\begin{equation}\begin{aligned}\label{eq:empirical_rule}
P(\mu_n - 1 \cdot \sigma_n \leq \kappa_n \leq \mu_n + 1 \cdot \sigma_n )&\approx 68.27\% \\
P(\mu_n - 2 \cdot \sigma_n \leq \kappa_n \leq \mu_n + 2 \cdot \sigma_n )&\approx 95.45\% \\
P(\mu_n - 3 \cdot \sigma_n \leq \kappa_n \leq \mu_n + 3 \cdot \sigma_n )&\approx 99.73\%
\end{aligned}\end{equation}

However, $\mu_n$ and $\sigma_n$ are largely biased by the subset $\mathcal{P}$ of $k$ participants selected in the first step of our experiment (cf. Table~\ref{table:mc}), because this influences the agreement value $\hat{\kappa}$ of $\mathcal{P}$. To nullify this effect, we chose to only compute $m = 100$ values this time and instead repeating the whole experiment a total of $j$ times, so that we gather data from $j$ different randomly selected development teams of size $k = 7$. We therefore updated the values for $\bar{\mu_n}$ and $\bar{\sigma_n}$ iteratively after each repetition $j$ of the Monte Carlo experiment:

\begin{equation}\begin{aligned}
\bar{\mu_n} = \frac{(j - 1) \cdot {\mu_n}_{j-1} + {\mu_n}_j}{j} \\
\bar{\sigma_n} = \frac{(j - 1) \cdot {\sigma_n}_{j-1} + {\sigma_n}_j}{j} 
\end{aligned}\end{equation}

Therefore, with numerous repetitions of $j$ the values of $\bar{\mu_n}$ and $\bar{\sigma_n}$ stabilize by regression to the mean, which should be the most representable values of $\bar{\mu_n}$ and $\bar{\sigma_n}$ for our data. To represent the resulting values meaningfully, we chose to determine the coefficient of variation $cv_n$~\cite{stepniak2011coeficientofvar}:

\begin{equation}\begin{aligned}
cv_n = \frac{\bar{\sigma_n}}{\bar{\mu_n}}
\end{aligned}\end{equation}

Assuming that $\bar{\mu_n}$ stays centered to $\hat{\kappa}$ even for small $n$, which we already know by answering RQ1, $cv_n$ therefore gives us the expected variation of $\kappa_n$. That is, $\approx 68.27\%$ of $\kappa_n$ are estimated in the interval 

\begin{align}\hat{\kappa} - cv_n \cdot \hat{\kappa} \leq \kappa_n \leq  \hat{\kappa} + cv_n \cdot \hat{\kappa}.
\end{align} 
In the same way, $\approx 95.45\%$ of $\kappa_n$ are estimated within the interval 
\begin{align}\hat{\kappa} - 2 \cdot cv_n \cdot \hat{\kappa} \leq \kappa_n \leq  \hat{\kappa} + 2 \cdot cv_n \cdot \hat{\kappa},
\end{align}
et cetera (cf. Equation~\ref{eq:empirical_rule}). By multiplying $cv_n \cdot 100$, we obtain the expected variation of these intervals in percentage relative to $\hat{\kappa}$.

\section{Results}\label{sec:results}
We performed our analysis as described in Section~\ref{subsec:analysis}. In the following, we present the results per research question. 

\subsection{Results for RQ1}
\begin{figure*}[tb]
	\begin{center}
	    \includegraphics[width=0.6666\textwidth]{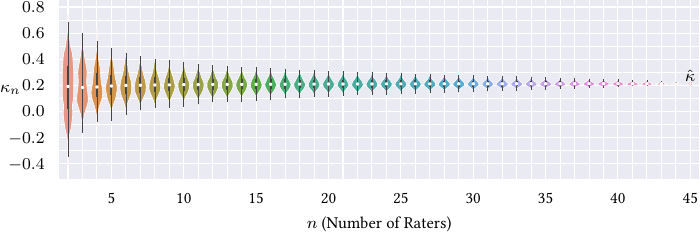}
	\end{center}
    \caption{Violin plot of the agreement values $\boldsymbol{\kappa_n}$ resulting from one run of the Monte Carlo experiment (cf. Table~\ref{table:mc}). Each violin and their inner box plot contain $\boldsymbol{m = 1000}$ agreement values $\boldsymbol{\kappa_n}$ corresponding to a subset of $\boldsymbol{n}$ out of $\boldsymbol{k = 45}$ participants.}
    \Description[The values of $\kappa_n$ increase in interquartile range for small $n$.]{With a decreasing number $n$ of raters  (i.e. small subsets of all 45 raters) the deviations of agreement ($\kappa_n$) expand both positively and negatively.}
\label{fig:mc_plot}
\end{figure*}

Using the Monte Carlo experiment described in Section~\ref{subsec:rq1}, we are able to compute numerous randomly generated progressions of deviating agreement for an increasing number of raters. Figure~\ref{fig:mc_plot} shows the resulting data of one run of the Monte Carlo experiment visualized for all $k = 45$ participants in a violin plot with inner box plots. The x-axis represents the number of unique raters $2 \leq n \leq k$ considered in each progressive step, while the y-axis shows the distribution of the $m = 1000$ calculated $\kappa_n$-values corresponding to each $n$. As visible, the interquartile range of $\kappa_n$ increases with smaller sized subsets $n$, as one would expect. This relationship is non-linear, meaning the smaller the subset of participants, the larger the deviations of $\kappa_n$ from $\hat{\kappa}$ can be expected. The median of the agreement values however stays centered around $\hat{\kappa}$ even for small $n$, meaning that positive and negative deviations of $\kappa_n$ from $\hat{\kappa}$ are equally alike. The data showed great uniformity, as depicted in Figure~\ref{fig:mc_plot}, with each repetition of the Monte Carlo experiment.

What is even more interesting, is that the same also holds true for smaller maximum group sizes of $k < 45$ raters. This is remarkable, since for $k = 45$ the final agreement $\hat{\kappa}$ will always be the same, but when selecting a subset of $k < 45$ raters $\mathcal{P}$ in the first step of Table~\ref{table:mc}, then $\hat{\kappa}$ will depend on $\mathcal{P}$ and differ for each repetition of the Monte Carlo experiment (cf.~\ref{subsec:rq1}). 

\subsection{Results for RQ2}
Answering RQ2 asks for a mathematical model that describes the minimum agreement $\min \kappa_n$ per number of raters $n$, as visualized in Figure~\ref{fig:mc_plot}. We took a closer look at the curve of the minima in dependency from the number of raters, and started finding a mathematical model that vaguely fitted our data. A promising candidate for our regression model was based on a rational function: 
\begin{equation}\begin{aligned}
	\min \kappa_n \approx f(n, a, b) = \frac{a \cdot n}{b + n} 
\end{aligned}\end{equation}

That means that we tried to approximate $\min \kappa_n$ with the model $f(n, a, b)$, which depends on the number of raters $n$ and two regressors $a$ and $b$. However, while $f$ followed the general shape of our real values for $\min \kappa_n$, the shape was offset in the x- and y-directions compared to our data. 
To compensate this offset, we added two more regressors $c$ and $d$ representing the vertical and horizontal shift of $f$ correspondingly:
\begin{equation}\begin{aligned}
	\Rightarrow f(n, a, b, c, d) = \frac{a \cdot (n - d)}{b + n - d} + c
\end{aligned}\end{equation}

Using this model, we performed a non-linear multiple regression analysis in \textit{SciPy}~\cite{virtanen2020scipy}. To enhance the results, we used an initial guess, which we derived by selecting values for $a, b, c$ and $d$ that fitted to our data in Figure~\ref{fig:mc_plot} by using a graphical calculator. This was necessary to compute actually fitting solutions for the parameter using the regression analysis with 4 regressors. We computed the regression repeatedly for each run of the Monte Carlo experiment as described in section~\ref{subsec:rq2}, using the resulting values for $\min \kappa_n$ of each subsequent run. Repeating this process multiple times, we noticed that the value of the regressor $d$ always settled close to a value of 2. This behavior can be explained by the fact that the $\min \kappa_n$ values  start at $n = 2$ raters, with no prior data (leading to a shift on the x-axis). Therefore, we replaced the regressor with a constant value of $d = 2$ in our model accordingly:
\begin{equation}\begin{aligned}
	\Rightarrow f(n, a, b, c) = \frac{a \cdot (n - 2)}{b + n - 2} + c
\end{aligned}\end{equation}

We repeated the regression analysis process using the improved formula with one less regressor to optimize. We found that the variable $c$ now stabilized to the value of $\hat{\kappa} - a$ (i.e., the final agreement value of the whole group of $k$ raters minus the regressor $a$). Therefore, we could further minimize the number of regressors by inserting the equation $c = \hat{\kappa} - a$ into $f$:

\begin{equation}\begin{aligned}
	f(n, a, b, \hat{\kappa}) = \frac{a \cdot (n - 2)}{b + n - 2} + \hat{\kappa} - a
\end{aligned}\end{equation}

Therefore, we used two known parameters $n$ and $\hat{\kappa}$ for our model $f$ with only two regressors $a$ and $b$ remaining. By repeating the regression analysis multiple times again with these two regressors, we found that $a$ always settled to a value of $2 \cdot \hat{\kappa}$ (i.e., double the value of the final agreement value $\hat{\kappa}$ of the whole group of $k$ raters). We could again replace one regressor of  our model by inserting the equation $a = 2 \cdot \hat{\kappa}$ into $f$:

\begin{equation}\begin{aligned}
    f(n, b, \hat{\kappa}) = \frac{2 \cdot \hat{\kappa} \cdot (n - 2)}{b + n - 2} - \hat{\kappa}
\end{aligned}\end{equation}

So far, the model fitted the real values of $\min \kappa$, independent of the final team size $k$. The last remaining regressor $b$ was seemingly the only regressor varying with differently sized teams of $k$ raters, out of which we model the minimum agreement $\min \kappa_n$ of a subset of $2 \leq n \leq k$ raters. Using various different sizes of $3 \leq k \leq 45$, we could deduce a linear relationship between the remaining regressor $b$ and $k$, where $b$ equals $\tfrac{k}{10}$. By inserting $b = \tfrac{k}{10}$ into $f$, we can therefore obtain our final regression model:

\begin{equation}\begin{aligned}
    f(n, k, \hat{\kappa}) = \frac{2 \cdot \hat{\kappa} \cdot (n -2)}{\frac{k}{10} + (n - 2)} - \hat{\kappa}
\end{aligned}\end{equation}

This is under the assumption that we contemplate a certain development team with $k$ members and an overall agreement of $\hat{\kappa}$. Under this assumption, $f(n, k, \hat{\kappa})$ returns the lower bound agreement value $\min \kappa_n$ for a random subset of $2 \leq n \leq k$ team members. For a more formal mathematical definition of our model, we suggest the following equation.

\begin{xdefinition}
\textbf{Theorem 1:} Let $\mathcal{R} = \lbrace r_i : 1 \leq i \leq k \rbrace$ be a team of $k$ raters. Let further $\hat{\kappa} = \kappa(\mathcal{R})$ be the overall agreement of the whole team. For a given $n$ with $1 < n < k$, we find that

\definecolor{shadecolor}{HTML}{f2f2f7} 
\begin{snugshade*}
\begin{equation*}
    {\displaystyle\min_{\mathcal{\kappa}_n}\lbrace\kappa_n\rbrace \; \approx \; \frac{2 \cdot \hat{\kappa} \cdot (n -2)}{\frac{k}{10} + n - 2} - \hat{\kappa}} 
\end{equation*}
 \end{snugshade*}

\noindent
where $\kappa_n = \kappa(\mathcal{R}_n)$ for $\mathcal{R}_n \subset \mathcal{R}$ with $|\mathcal{R}_n| = n$. 
\end{xdefinition}

\begin{figure*}[tb]
	\begin{center}
		\includegraphics[width=0.6666\textwidth]{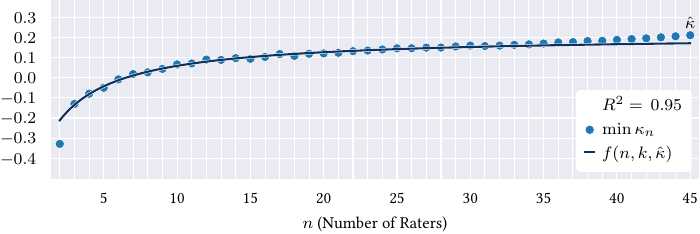}
	\end{center}
	\caption{Comparison of the minimum agreement $\boldsymbol{\min \kappa_n}$ from the Monte Carlo experiment and our deducted model $\boldsymbol{f(n, k, \hat{\kappa})}$ used for regression.}
    \Description[The regression model fits the actual agreement data.]{Our regression model $f(n, k, \hat{\kappa})$ is able to achieve high $R^2$-values compared to the minimum agreement $\min \kappa_n$ simulated from our data.}
	\label{fig:regression}
\end{figure*}

However, we want to point out that we cannot confirm the general applicability of this model, and neither prove its applicability formally at this point in time. Though for our data, this model generally produces results fitting the minimum agreement values $\min \kappa_n$ with a $R^2$-value of $0.9$ and above. Lower $R^2$-values could be explained by the randomly generated plot having a flatter curvature because more agreeing raters were selected in the beginning by chance of the Monte Carlo experiment. However, this happened only rarely. A comparison of our model $f(n, k, \hat{\kappa})$ with the actual lowest data points $\min \kappa_n$ for one run of the Monte Carlo experiment is shown in Figure~\ref{fig:regression}. 

Figure~\ref{fig:regression} shows how our regression model $f(n, k, \hat{\kappa})$  provides a close fit to the actual values of $\min \kappa_n$. For numbers of raters $n$ close to the total number of raters $k$, the model over-estimates the badness of $\min \kappa_n$. Although the model fitment is great for the real-world data used as a basis of our Monte Carlo experiment, it is obvious that for a theoretically perfectly agreeing team (i.e., $\hat{\kappa} = 1$), it does not matter how many or which raters of this team are removed, the resulting subset of $2 \leq n \leq k$ raters will always have the same agreement of $\kappa_n = 1 = \hat{\kappa}$. This is not reflected in our model, but is more of a thought experiment anyway as high agreement values especially for larger teams are unlikely by nature. 

\subsection{Results for RQ3}

We can now concretize the results for a team consisting of 7 team members. Table~\ref{table:cvn} shows the coefficients of variation $cv_n$ in dependency of $n$, as resulting from the methodology described in Section~\ref{subsec:rq3} ($k = 7,\; m = j = 100$). 

\begin{table}[ht]
\setlength{\tabcolsep}{4pt} % to further tweak if needed
\renewcommand{\arraystretch}{1.1} % adjust row height
\caption{Coefficient of variation $\boldsymbol{cv_n}$ for $\boldsymbol{n \in \{ 2, \ldots, 7 \}}$.} 
\begin{center}
\begin{tabularx}{\columnwidth}{@{}l*{6}{>{\centering\arraybackslash}X}@{}}
\Xhline{2\arrayrulewidth}
\multirow{2.25}{*}{$cv_n$} & $n=2$ & $n=3$ & $n=4$ & $n=5$ & $n=6$ & $n=7$ \\
\cmidrule(lr){2-2}
\cmidrule(lr){3-3}
\cmidrule(lr){4-4}
\cmidrule(lr){5-5}
\cmidrule(lr){6-6}
\cmidrule(lr){7-7}
& 1.1868 & 0.6549 & 0.4429 & 0.3050 & 0.1909 & 0.0000 \\
\Xhline{2\arrayrulewidth}
\end{tabularx} 
\label{table:cvn}
\end{center}
\end{table}

As expected, if all 7 members of a random development team are accounted for in the labeling of statements, no variation of their agreement occurs, since $\kappa_7 = \hat{\kappa}$ and therefore $cv_7 = 0$. More interestingly though, if we only exclude one random member from the 7 developers $cv_n$ increases rapidly. We observed $cv_6 = 0.1909$, meaning that the $\approx 68.27\%$ interval estimate of $\kappa_6$ is $\hat{\kappa} \pm 19.09\%$. In the same way, we observe an $\approx 68.27\%$ interval estimate of $\hat{\kappa} \pm 30.50\%$ for $\kappa_5$. As already visualized by Figure~\ref{fig:mc_plot}, $cv_n$ grows faster than linear with descending $n$ (cf. Table~\ref{table:cvn}). By only accounting for 2 raters out of 7, we observe the $\approx 68.27\%$ interval estimate of $\kappa_2$ to be $\hat{\kappa} \pm 118.68\%$. Accordingly, $\approx 68.27\%$ of $\kappa_n$ are expected within $\hat{\kappa} - 1.1868 \cdot \hat{\kappa} \leq \kappa_n \leq  \hat{\kappa} + 1.1868 \cdot \hat{\kappa}$. For our data in Table~\ref{table:cvn} we observed $\hat{\kappa} = 0.2193$.

\section{Discussion}\label{sec:discussion}
We end this paper by answering the research questions, interpreting our results, providing an outlook for future work, and discussing threats to validity. 

\subsection{Answers to Research Questions}
Based on our data analysis and the results presented in prior sections, we come to the following answers for each research question:

\begin{itemize}[leftmargin=25pt]
\setlength\itemsep{1mm}
    \item[\textbf{RQ1:}] \textit{How does the agreement of sentiment towards software project communication among a subset of developers change with decreasing subset size?} By computing numerous random possibilities using our Monte Carlo experiment, we can conclude that the volatility of agreement (i.e., Fleiss' $\kappa$) increases rapidly when drawing a smaller number of raters out of a bigger group. Therefore, the fewer raters are accounted for, the larger the deviation from the actual agreement of the whole group needs to be expected. The mean agreement values for smaller subsets of raters stay centered on the agreement of the whole group throughout this process, even though the confidence interval will increase. 
    
    \item[\textbf{RQ2:}] \textit{How can the minimum agreement for small developer subsets be mathematically modeled based on the overall sentiment agreement of the entire group?} We provided a mathematical model that approximates our data, but cannot generalize the model for all agreement data yet. More research is needed to evaluate this formula on different datasets of raters. If the formula is proven to be reliable across multiple datasets, it provides a way to estimate the minimum agreement boundary (this is under the assumption that the total number of raters as well as the agreement of the whole group are familiar).
    
    \item[\textbf{RQ3:}] \textit{What is the minimum number of developers required to achieve the same level of sentiment agreement as an average-sized software development team?} For a random development team with 7 members, which is a common size according to past and present literature, we have demonstrated which agreement deviations can be expected for a random subset of $2 \leq n \leq 7$ members by calculating the coefficient of variation regarding the $\approx 68.27\%$ interval estimate of agreement. We observed that even with 6 out of 7 members, this interval stretches to $\hat{\kappa} \pm 19.09\%$. These intervals increase rapidly for even smaller numbers of members, therefore becoming completely unreliable. We recommend, if one is serious about creating a robust dataset that reflects the perceptions of a specific development team, to not neglect any member of that team when creating the dataset.
\end{itemize}

\subsection{Interpretation}
While higher and lower agreement values are equally alike when considering the mood of a small subset of developers, we interpret the uncertainty about the actual outcome as a risk for unreliable mood surveying results within software development teams. Our findings do not necessarily contradict those of Murgia et al.~\cite{Murgia.2014}, who reported that the agreement did not change substantially beyond the first two raters. However, the effect observed by Murgia et al.~\cite{Murgia.2014} may rather be the result of concrete and well-defined coding guidelines, while we chose to observe the subjective perceptions of developers towards software project communication. As the violin plots within Figure~\ref{fig:mc_plot} revealed, agreement values close to the agreement of the whole group are still most probable for smaller subsets. However, these probabilities shrink and become more evenly distributed across larger intervals for a smaller subset of developers. 

Our findings have implications for collaboration within software development teams and offer valuable insights for project managers. Understanding the diversity of perceptions among team members is crucial, as it directly influences communication dynamics and can affect team morale and project success. By recognizing the varied ways in which individual developers interpret communication, project managers can better address the emotional subjectivity that leads to differing sentiment interpretations. This awareness allows for the early identification of potential miscommunications, fostering a more harmonious and effective collaborative environment. However, conducting sentiment analysis on developer communication raises ethical concerns such as privacy, consent, and potential self-censorship~\cite{specht2024hcse}. Without explicit consent, analyzing developers' messages intrudes on privacy and may lead to altered, inauthentic communication. If developers know their interactions are being analyzed, they might change their tone, hindering open collaboration. Ethical approaches must ensure that sentiment analysis supports an authentic and respectful communicative environment.

Several influential factors can affect the expression and perception of sentiments in communication, and therefore explain our observations. Different developers have different perceptions of communication~\cite{herrmann2025different-perceptions}. Human values, such as those outlined in Schwartz's theory of basic human values~\cite{schwartz1992values}, which identifies ten fundamental values and can be measured using surveys like the Short Schwartz's Value Survey~\cite{lindeman2005ssvs}, could significantly impact sentiment perception in software projects~\cite{winter2018measuring, liebel2024human-mde, wohlrab2025human-values}. Additionally, personality traits, as modeled by the OCEAN framework, including \textit{openness}, \textit{conscientiousness}, \textit{extraversion}, \textit{agreeableness}, and \textit{neuroticism}~\cite{madison2021ocean}, may influence communication perceptions in software projects~\cite{vishnubhotla2020personalities}. Furthermore, team structure and project nature may influence the general communication style on an organizational level.

Further, we examined the diversity of perceptions within development teams and sought to determine how many team members need to participate in sentiment reporting to accurately capture the team's mood. Through a Monte Carlo experiment involving 45 developers, we proposed a preliminary mathematical model to estimate the minimum agreement necessary among a subset of raters to align with the entire group’s consensus. Our findings suggest that excluding even a single member from mood assessments in an average team can lead to inaccuracies in representing the overall mood. Consequently, for project managers, this means that comprehensive sentiment data can facilitate more informed decision-making regarding team dynamics and project strategies. With the insights gained from our study, managers can implement tailored surveying strategies that align with the specific perceptions of their teams, enabling real-time mood monitoring and more precise interventions when necessary. This ensures that mood surveying results remain aligned with the team’s shifting perceptions. This adaptability can lead to more reliable mood assessments and, ultimately, to a more cohesive team atmosphere.

Despite advances in sentiment analysis, existing off-the-shelf tools and datasets for SA4SE are not adequate for precise mood evaluations without adjustments tailored to specific teams. Creating datasets that accurately reflect the team’s mood necessitates the involvement of all team members, as neglecting even one person undermines dataset reliability. Annotating sufficiently large datasets poses a challenge due to the effort required. Therefore, developing scalable procedures, such as semi-automated labeling, is vital. This approach allows sentiment analysis tools to be dynamically aligned with the varying perceptions of team members, thereby improving the accuracy of mood assessments relative to the subjective perspectives within the development team.

\subsection{Future Work}
Building upon the insights gained in this study, future work should focus on further refining mood analysis approaches to better accommodate the diverse perceptions present within software development teams. One area for future investigation is the development of best practice frameworks for mood surveying in software environments. Such frameworks would consider factors like team size, diversity, and communication culture to optimize mood data collection strategies. Researchers could also explore the impact of various team interventions, on improving team cohesion and project outcomes. Furthermore, longitudinal studies tracking sentiment perception changes over project lifecycles could yield deeper insights into how team dynamics evolve and influence project success. This direction not only enhances the predictive power of sentiment analysis tools but also supports the creation of more effective management strategies in software development.   

Additionally, studies that track the mood and communication patterns of development teams over time would provide deeper insights into how sentiment analysis can anticipate and mitigate potential misunderstandings. Future research should also explore the integration of these tailored sentiment tools into project management workflows, examining how real-time mood assessment directly impacts project outcomes and team cohesion. By translating these findings into practical applications, project managers can be better equipped to foster a collaborative environment that proactively addresses communication barriers and enhances overall team performance.

\subsection{Threats To Validity}
While this paper focuses on the analysis of agreement in teams of software developers, our research methodology is generally applicable for analyzing the behavior of multiple human raters in subjective labeling tasks. However, different results are to be expected. This paper is subject to the same threats caused by the origin of the data that were presented in Section~5.4 of our preceding paper~\cite{herrmann2022subjectivity}. In the following, we will reiterate these threats and add threats resulting from the methodology of this paper. We classify the potential threats to our approach analysis based on the categorization proposed by Wohlin et al.~\cite{Wohlin.2012} as construct, internal, conclusion, and external validity.

\subsubsection{Construct Validity}
{Language Proficiency}. As the researchers were located in a specific region, the majority of our population consisted of non-native English speakers. However, the statements from the datasets were entirely in English, including technical terms. To address this concern, we inquired about participants' frequency of communication in English. Approximately two-thirds of our population reported communicating in English at least once a week, while one-third communicated in English less frequently. Nevertheless, we assume that all participants considered their English comprehension sufficient for completing the survey. Furthermore, it should be noted that most technical terms in the software engineering and computer science domain are in English, regardless of the language used. 

{Survey Environment}. Since the survey was conducted at participants' homes, there may have been potential distractions and interferences for individual participants. {Potential Overlap of Raters}. The study was also distributed among colleagues, with a request to further distribute it to other potential candidates. Therefore, it is possible that raters of the two datasets used in our paper also participated. While we could not exclude this in advance, we believe the impact is negligible due to the large number of participants involved.

\subsubsection{Internal Validity}
{Selection Bias}. For our analysis procedures, we only accounted for 45 raters who had labeled all the 100 statements present in the survey~\cite{sentisurvey-zenodo}, instead of the 94 raters who had at least annotated one statement each~\cite{herrmann2022subjectivity}. By applying a complete case analysis, we therefore may have induced a bias if the annotations of the removed participants were not missing completely at random (MCAR~\cite{rubin1976mcar-mar-mnar}). If the annotations were instead missing at random (MAR~\cite{rubin1976mcar-mar-mnar}) the absence of a participant's annotations would correlate with an independent variable such as work experience. In this case, it would follow that we create a bias that favors study participants with more work experience.

\subsubsection{Conclusion Validity}
{Lack of Context}. Presenting isolated statements without conversational context might have affected participants' ability to accurately assign sentiment polarity. While intended to capture initial emotional reactions, this approach limits the capture of nuanced sentiment. 

{Computational Limitations}. Despite the thoroughness of our Monte Carlo methodology, it does not account for every possible combinations (i.e., $\ll 44! \approx 2.7 \cdot 10^{54}$) due to computational constraints. This limitation necessitates cautious interpretation of predictive power.

\subsubsection{External Validity}
{Monte Carlo Simulation}. The results of our study are based on simulations. To enhance the practical reliability and validity our results need to be confirmed by real-world experiments.

{Population Specificity}. The majority of participants were student software developer with little professional experience, limiting the generalizability of findings. While this is often the case in dataset creation (cf.~\cite{novielli2018stackgold}), future research should further examine group-specific behaviors to ensure broader applicability. 

{Representativeness}. Our sample included participants with relevant programming experience, supporting confidence that results reflect perceptions typical within the software engineering domain.

\section{Conclusion}\label{sec:conclusion}
In this study, we employed a Monte Carlo experiment to investigate the variability in agreement among subsets of raters compared to the entire group. Our findings demonstrate that significant deviations in sentiment agreement are likely within smaller subsets, underscoring the importance of involving all team members to achieve a comprehensive and accurate assessment when surveying the teams' mood. We developed a preliminary mathematical model to calculate the minimum agreement needed for subsets based on the group's overall consensus. This has practical implications for mood surveying, as for an average sized team of 7 developers, our results point to the necessity of surveying all team members. Excluding even one individual can skew the overall mood assessment, potentially leading to misunderstandings and disruptions in collaborative efforts. 

Further, our findings underscore the criticality of comprehensive sentiment analysis in enhancing team dynamics and communication, such as the need for care and precision in constructing datasets for sentiment analysis for software engineering. 

As we move forward, developing best practice frameworks for mood surveying in agile software environments that consider team size, diversity, and communication culture can optimize sentiment data collection strategies, while exploring the impact of team interventions and conducting longitudinal studies on sentiment perception changes can lead to enhanced predictive power of sentiment analysis tools and more effective management strategies in software development.

%\begin{acks}
%Omitted for anonymity.
%This research was funded by the Leibniz University Hannover as a Leibniz Young Investigator Grant (Project \textit{ComContA}, Project Number \textit{85430128}, 2020--2022).
%\end{acks}


\section*{Data Availability}
The raw dataset containing the unprocessed survey results and associated metadata is available on Zenodo under the following DOI:~\href{https://doi.org/10.5281/zenodo.6611728}{10.5281/zenodo.6611728}.

\balance
\bibliographystyle{ACM-Reference-Format}
\bibliography{references}

\end{document}